\documentclass[12pt]{article}
\begin{document}
\title{Energy and Mass Generation}
\author{B.G. Sidharth\\
International Institute for Applicable Mathematics \& Information Sciences\\
Hyderabad (India) \& Udine (Italy)\\
B.M. Birla Science Centre, Adarsh Nagar, Hyderabad - 500 063
(India)}
\date{}
\maketitle
\begin{abstract}
Modifications in the energy momentum dispersion laws due to a
noncommutative geometry, have been considered in recent years. We
examine the oscillations of extended objects in this perspective and
find that there is now a "generation" of energy.
\end{abstract}
\section{Introduction}
Many modern approaches, particularly in Quantum Gravity introduce a
minimum spacetime scale. It is well known that the introduction of
such a fundamental minimum length in the universe leads to a
noncommutative geometry. On the other hand there has been much
discussion about how this could avert the infinities which plague
Quantum Field Theory (Cf.ref.\cite{tduniv} and several other
references therein). As pointed out by Snyder many years ago
\cite{snyder}, there are now new commutation relations which replace
the usual Quantum Mechanical relations. These are
$$[x,y] = (\imath l^2/\hbar )L_z, [t,x] = (\imath l^2/\hbar c)M_x, etc.$$
\begin{equation}
[x,p_x] = \imath \hbar [1 + (l/\hbar )^2 p^2_x];\label{2De3}
\end{equation}
where $a$ is the fundamental length.\\
It was noted by the author that these new commutation relations lead
to a modified energy momentum relation viz.,
\cite{tduniv,kg,bgsijtp2004}
\begin{equation}
E^2 = p^2+m^2+\alpha l^2 p^4\label{2}
\end{equation}
giving the so called Snyder-Sidharth Hamiltonian \cite{glinka},
$\alpha$ being a scalar, and $l$ the fundamental length, in units $c
= 1 = \hbar$. This leads to a modification of the usual Klein-Gordon
equation, which now becomes,
\begin{equation}
(D + l^2 \nabla^4 - m^2)\psi = 0\label{6.52}
\end{equation}
where $D$ denotes the usual D'Alembertian.\\
It is well known that the usual Klein-Gordon equation, without the
extra term in (\ref{2}) describes a super position of normal mode
harmonic oscillators \cite{bd2}\\
Let us first start with the Klein-Gordon equation itself
\begin{equation}
(D + m^2) \psi (x) = 0\label{12.1}
\end{equation}
Following a well known procedure, a Fourier integral over elementary
plane waves is then taken to give
$$\psi (x,t) = \int \frac{d^3k}{\sqrt{(2\pi)^3 2_{\omega k}}} \left[a
(k)e^{\imath k.x-\imath \omega_k t} + a^\dag (k) e^{-\imath
k.x+\imath \omega_k t}\right]$$
\begin{equation}
= \int d^3k \left[a (k) f_k (x) + a^\dag (k) f^*_k
(x)\right]\label{12.7}
\end{equation}
where
$$\omega_k = + \sqrt{k^2 + m^2} \, \mbox{and} \, f_k (x) =
\frac{1}{\sqrt{(2 \pi)^3 2 \omega_k}} e^{-\imath k.x}$$ An inversion
then gives
$$\int f^*_k (x,t) \psi (x,t)d^3 x = \frac{1}{2 \omega_k} \left[a (k) + a^\dag
(-k) e^{2\imath \omega_k t}\right]$$
\begin{equation}
\int f^*_k (x,t) \dot{\psi} (x,t) d^3 x = \frac{- \imath}{2} \left[a
(k) - a^\dag (-k) e^{2\imath \omega_k t}\right]\label{12.8}
\end{equation}
Finally the canonical Quantum Mechanical commutation relations for
the wave function (now operator) $\phi$ and the associated momentum
leads to (Cf.ref.\cite{bd2} for details)
$$[a(k), a^\dag (k')] = \int d^3 x d^3 y [f^*_k (x,t) \partial_0
\psi (x,t), f k' (y,t) \partial_0 \psi (y,t)]$$
$$= + \imath \int d^3 x f^*_k (x,t) \partial_0 fk' (x,t) = \delta^3
({\bf k - k'})$$
Similarly
$$[a(k),a(k')] = (\imath)^2 \int d^3x d^3y [f^*_k (x,t) \partial_0
\psi (x,t), f^*_{k'} (x,t) = 0$$ and
\begin{equation}
[a^\dag (k), a^\dag (k')] = 0\label{12.10}
\end{equation}
This finally gives the Hamiltonian in terms of the creation and
annihilation operators $a$ and $a^\dagger$ viz.,
\begin{equation}
H = 1/2 \int d^3k \omega_k [a^\dag (k) a(k) + a(k) a^\dag
(k)]\label{12.11}
\end{equation}
Discretizing the above, replacing the integrals in momentum space by
summatiions and the Dirac-Delta function by the Kronecker Delta, we
get, as is well known,
\begin{equation}
H = \sum_{k} H_k = \sum_{k} 1/2 \omega_k (a^\dag_k a_k + a_k
a^\dag_k) \quad a_k = \sqrt{\Delta V_k} a(k)\label{12.13}
\end{equation}
with
\begin{equation}
[a_k, a^\dag_{k'}] = \delta_{k k'} \quad [a_k,a_{k'}] = [a^\dag_k,
a^\dag_{k'}] = 0\label{12.14}
\end{equation}
In this second quantized picture, these lead to the various particle
solutions given by
\begin{equation}
H_k \Phi_k (n_k) = \omega_k (n_k + 1/2) \Phi_k (n_k)\label{12.15}
\end{equation}
\section{The New Scenario}
Let us see, how all this gets modified due to the new considerations
and in particular the new commutation relations (\ref{2De3}). We
first observe that if instead of the usual energy momentum relation,
we use (\ref{2}) above, then this would lead to a modification of
the frequency given in (\ref{12.7}). That is we would have
\begin{equation}
\omega_k = \sqrt{k^2+m^2+\alpha l^2 k^4} \equiv \omega'_k\label{A}
\end{equation}
(Remembering that we are in natural units). This would mean that
there would be a shift in the energy of the individual oscillators
(Cf. also \cite{bgsijmpe2010,artg}). But also, there are other important changes.\\
To see this, let us consider the above in greater detail. As is well
known for the individual Quantum Harmonic oscillators in
(\ref{12.13}) or (\ref{12.15}) we have \cite{bd2}
$$a = \lambda_1 q + \imath \lambda_2 p$$
\begin{equation}
a^\dag = \lambda_1 q - \imath \lambda_2 p\label{X}
\end{equation}
where $\lambda_1 = \left(\frac{m\omega}{2}\right)^{\frac{1}{2}}$ and
$\lambda_2 = \left(\frac{1}{2m\omega}\right)^{\frac{1}{2}}$, $q$
replaces $x$ and we have dropped the subscript $k$ for $\omega$.
This leads to, in the old case the Hamiltonian
\begin{equation}
H_0 = \omega (a^\dag a + \frac{1}{2})\label{Y}
\end{equation}
In our case however there is the additional term in (\ref{2}). To
see this in greater detail we note that owing to the commutation
relations (\ref{2De3}) we have
\begin{equation}
[q,p] = \imath (1 + l^2 p^2)\label{a2}
\end{equation}
and so,
\begin{equation}
[a, a^\dagger ] = (1 + l^2 p^2)\label{a3}
\end{equation}
Whence we have
$$H = \frac{m \omega}{2} [a a^\dagger + a^\dagger a]$$
$$= \frac{m \omega}{2} [2a^\dagger a + (1 + l^2 p^2)]$$
\begin{equation}
= m \omega [a^\dagger a + \frac{1}{2} (1 + l^2 p^2)]\label{Ba}
\end{equation}
Finally we have
\begin{equation}
H = m \omega [(a^\dagger a + \frac{1}{2}) + \frac{l^2}{2}
p^2]\label{b1}
\end{equation}
$$= H'_0 + \frac{m \omega}{2} l^2 p^2$$
where $H'_0$ is given by
\begin{equation}
H'_0 \equiv m \omega (a^\dagger a + \frac{1}{2})\label{c1}
\end{equation}
that is, it would be the Hamiltonian if $l^2$ were $0$.\\
Using (\ref{X}) it is easy to see that $p^2$ is given by
\begin{equation}
p^2 = (\frac{m \omega}{2}) (a - a^\dagger )^2\label{c3}
\end{equation}
Using (\ref{c3}) in (\ref{b1}), we get finally

\begin{equation}
H = H'_0 + l^2 H \frac{m \omega}{2} - (\frac{m \omega l}{2})^2 (a^2
+ a^{\dagger 2})\label{d}
\end{equation}
Equation (\ref{d}) for the Quantum Mechanical Harmonic Oscillator
already shows the effect of the extra term in the SS-Hamiltonian
(\ref{2}). This is an additional energy that appears due to the
commutation relations (\ref{2De3}).\\
But it also shows that apart from the shift in energy (or mass) for
the individual oscillators, reflecting (\ref{2}), the eigen states
of the oscillators get mixed up due to the presence of the squares
of the creation and annihilation operators. The eigen states would
be combinations of states like $\Phi (n-2), \Phi (n+2)$ in addition
to $\Phi (n)$, of the usual theory as in (\ref{12.15}). In other
words we have to consider a system of oscillators. This is due to
the fact that a point is being replaced by an extension.\\
Indeed in String Theory itself, as we know we have a similar
situation \cite{tduniv} and \cite{fuzzy}-\cite{jacob}. We have
\begin{equation}
\rho \ddot {y} - T y'' = 0,\label{ce1}
\end{equation}
\begin{equation}
\omega = \frac{\pi}{2l} \sqrt{\frac{T}{\rho}},\label{ce2}
\end{equation}
\begin{equation}
T = \frac{mc^2}{l}; \quad \rho = \frac{m}{l},\label{ce3}
\end{equation}
\begin{equation}
\sqrt{T/\rho} = c,\label{ce4}
\end{equation}
$T$ being the tension of the string, $l$ its length and $\rho$ the
line density and $\omega$ in (\ref{ce2}) the frequency. The
identification (\ref{ce2}),(\ref{ce3}) gives (\ref{ce4}), where $c$
is the velocity of light, and (\ref{ce1}) then goes over to the
usual d'Alembertian or massless Klein-Gordon equation and an expansion in normal
modes (which are uncoupled).\\
Further, if the above string is quantized canonically, we get
\begin{equation}
\langle \Delta x^2 \rangle \sim l^2.\label{ce5}
\end{equation}
Thus the string can be considered as an infinite collection of
harmonic oscillators \cite{fog}. Further we can see, using equations
(\ref{ce2}) and (\ref{ce3}) and the fact that
$$\hbar \omega = mc^2$$
and that the extension $l$ is of the order of the Compton wavelength
in (\ref{ce5}), a circumstance that was called one of the miracles
of the string theory by Veneziano \cite{ven,ven2,ven3,ven4,ven5}.\\
The point is that a non zero extension for the system of oscillators
would imply the commutation relations (\ref{2De3}). This again
implies an extra energy or mass on the one hand, and a coupled
system of oscillators on the other, rather than a superposition of
normal modes as we would otherwise have. Thus a consideration of a
system of coherent oscillators would be a better starting point, if
we consider the Planck scale, as in string theory and other
approaches, rather than spacetime points of the usual theory. This
circumstance has been
overlooked.\\
In summary there is now an extra energy or mass that appears due to
(\ref{2De3}) and (\ref{2}), on the one hand. On the other, we have
to consider a system of (coupled) oscillators.
\section{A system of Oscillators}
We know that String Theory, Loop Quantum Gravity, the author's own
approach of Planck oscillators, and a few other approaches start
from the Planck scale.\\
We will consider the problem from a different point of view, which
also enables an elegant extension to the case of the entire
\index{Universe}Universe itself \cite{bgspkos,uof,bhtd,emgphys}. Let
us consider an array of $N$ particles, spaced a distance $\Delta x$
apart, which behave like oscillators that are connected by springs.
This is because of the results in the last section that we have to
consider coupled oscillators. The starting point for such a
programme can be found in \cite{altaisky}.We then have
\cite{bgsfpl152002,good,vandam,uof}
\begin{equation}
r  = \sqrt{N \Delta x^2}\label{4De1d}
\end{equation}
\begin{equation}
ka^2 \equiv k \Delta x^2 = \frac{1}{2}  k_B T\label{4De2d}
\end{equation}
where $k_B$ is the Boltzmann constant, $T$ the temperature, $r$ the
extent  and $k$ is the spring constant given by
\begin{equation}
\omega_0^2 = \frac{k}{m}\label{4De3d}
\end{equation}
\begin{equation}
\omega = \left(\frac{k}{m}a^2\right)^{\frac{1}{2}} \frac{1}{r} =
\omega_0 \frac{a}{r}\label{4De4d}
\end{equation}
We now identify the particles with \index{Planck}Planck
\index{mass}masses and set $\Delta x \equiv a = l_P$, the
\index{Planck}Planck length. It may be immediately observed that use
of (\ref{4De3d}) and (\ref{4De2d}) gives $k_B T \sim m_P c^2$, which
of course agrees with the temperature of a \index{black hole}black
hole of \index{Planck}Planck \index{mass}mass. Indeed, Rosen
\cite{rosen} had shown that a \index{Planck}Planck \index{mass}mass
particle at the \index{Planck scale}Planck scale can be considered
to be a \index{Universe}Universe in itself a Schwarzchild Black Hole
of radius equalling the Planck length. We also use the fact deduced
earlier to that a typical elementary particle like the
\index{pion}pion can be considered to be the result of $n \sim
10^{40}$ \index{Planck}Planck
\index{mass}masses.\\
Using this in (\ref{4De1d}), we get $r \sim l$, the \index{pion}pion
\index{Compton wavelength}Compton wavelength as required. Whence the
pion mass is given by
$$m = m_P/\sqrt{n}$$
Further, in this latter case, using (\ref{4De1d}) and the fact that
$N = n \sim 10^{40}$, and (\ref{4De2d}),i.e. $k_BT = kl^2/N$ and
(\ref{4De3d}) and (\ref{4De4d}), we get for a \index{pion}pion,
remembering that $m^2_P/n = m^2,$
$$k_ B T = \frac{m^3 c^4 l^2}{\hbar^2} = mc^2,$$
which of course is the well known formula for the Hagedorn
temperature for \index{elementary particles}elementary particles
like \index{pion}pions \cite{sivaramamj83}. In other words, this
confirms the earlier conclusions that we can treat an elementary
particle as a series of some $10^{40}$ \index{Planck}Planck
\index{mass}mass
oscillators.\\
However it must be observed from (\ref{4De4d}) and (\ref{4De3d}),
that while the \index{Planck}Planck \index{mass}mass gives the
highest energy state, an elementary particle like the
\index{pion}pion is in the lowest energy state. This explains why we
encounter \index{elementary particles}elementary particles, rather
than \index{Planck}Planck \index{mass}mass particles in nature.
Infact as already noted \cite{bhtd}, a \index{Planck}Planck
\index{mass}mass particle decays via the \index{Bekenstein
radiation}Bekenstein radiation within a \index{Planck time}Planck
time $\sim 10^{-42}secs$. On the other hand, the lifetime of an
elementary particle
would be very much higher.\\
In any case the efficacy of our above oscillator model can be seen
by the fact that we recover correctly the \index{mass}masses and
\index{Compton scale}Compton scales in the order of magnitude sense
and also get the correct Bekenstein and Hagedorn formulas as seen
above, and further we even get the correct estimate of the
\index{mass}mass and size of the
\index{Universe}Universe itself, as will be seen below.\\
Using the fact that the \index{Universe}Universe consists of $N \sim
10^{80}$ \index{elementary particles}elementary particles like the
\index{pion}pions, the question is, can we think of the
\index{Universe}Universe as a collection of $n N \, \mbox{or}\,
10^{120}$ Planck \index{mass}mass oscillators? This is what we will
now show. Infact if we use equation (\ref{4De1d}) with
$$\bar N \sim 10^{120},$$
we can see that the extent is given by $r \sim 10^{28}cms$ which is
of the order of the diameter of the \index{Universe}Universe itself.
We shall shortly justify the value for $\bar{N}$. Next using
(\ref{4De4d}) we get
\begin{equation}
\hbar \omega_0^{(min)} \langle \frac{l_P}{10^{28}} \rangle^{-1}
\approx m_P c^2 \times 10^{60} \approx Mc^2\label{4De5d}
\end{equation}
which gives the correct \index{mass}mass $M$, of the
\index{Universe}Universe which in contrast to the earlier
\index{pion}pion case, is the highest energy state while the
\index{Planck}Planck oscillators individually are this time the
lowest in this description. In other words the
\index{Universe}Universe itself can be considered to be described in
terms of normal modes of
\index{Planck scale}Planck scale oscillators.\\
More generally, if an arbitrary mass $M$, as in (\ref{4De5d}), is
given in terms of $\bar{N}$ Planck oscillators, in the above model,
then we have from (\ref{4De5d}) and (\ref{4De1d}):
$$M = \sqrt{\bar{N}} m_P \, \mbox{and}\, R = \sqrt{\bar{N}} l_P,$$
where $R$ is the radius of the object. Using the fact that $l_P$ is
the Schwarzchild radius of the mass $m_P$, this gives immediately,
$$R = 2GM/c^2$$
a relation that has been deduced alternatively. In other words, such
an object, the Universe included as a special case, shows up as a
Black Hole, a super massive Black Hole in this case.
\section{Discussion}
The interesting question is, from where does the extra energy come?
We must remember that in the above considerations we have an a
priori dark energy background, and it is this energy that manifests
itself as the extra energy or mass or frequency.\\
This can be illustrated by considering the background
electromagnetic field, or the Zero Point Field as a collection of
ground state oscillators (Cf.ref.\cite{tduniv} and references
therein). It is known that the probability amplitude is
$$\psi (x) = \left(\frac{m\omega}{\pi \hbar}\right)^{1/4} e^{-(m\omega/2\hbar)x^2}$$
for displacement by the distance $x$ from its position of classical
equilibrium. So the oscillator fluctuates over an interval
$$\Delta x \sim (\hbar/m\omega)^{1/2}$$
The background \index{electromagnetic}electromagnetic field is an
infinite collection of independent oscillators, with amplitudes
$X_1,X_2$ etc. The probability for the various oscillators to have
amplitudes $X_1, X_2$ and so on is the product of individual
oscillator amplitudes:
$$\psi (X_1,X_2,\cdots ) = exp [-(X^2_1 + X^2_2 + \cdots)]$$
wherein there would be a suitable normalization factor. This
expression gives the probability amplitude $\psi$ for a
configuration $B (x,y,z)$ of the magnetic field that is described by
the Fourier coefficients $X_1,X_2,\cdots$ or directly in terms of
the magnetic field configuration itself by, as we saw,
$$\psi (B(x,y,z)) = P exp \left(-\int \int \frac{\bf{B}(x_1)\cdot \bf{B}(x_2)}{16\pi^3\hbar cr^2_{12}} d^3x_1 d^3x_2\right).$$
$P$ being a normalization factor. At this stage, we are thinking in
terms of energy without differenciation, that is, without
considering Electromagnetism or Gravitation etc as separate. Let us
consider a configuration where the magnetic field is everywhere zero
except in a region of dimension $l$, where it is of the order of
$\sim \Delta B$. The probability amplitude for this configuration
would be proportional to
$$\exp [-((\Delta B)^2 l^4/\hbar c)]$$
So the energy of \index{fluctuation}fluctuation in a region of
length $l$ is given by finally, the density \cite{mwt,uof}
$$B^2 \sim \frac{\hbar c}{l^4}$$
So the energy content in a region of volume $l^3$ is given by
\begin{equation}
\beta^2 \sim \hbar c/l\label{4e1}
\end{equation}
Equation (\ref{4e1}) can be written as
\begin{equation}
\mbox{Energy}\, \sim m_ec^2 \sim \frac{\hbar c}{l}\label{Y}
\end{equation}
Equation (\ref{Y}) shows that $l$ is the Compton wavelength of an
elementary particle, which thus arises naturally. On the other hand
if in (\ref{Y}), we consider the energy of a Planck mass rather than
that of an elementary particle, we will get the Planck length.\\
For another perspective, it is interesting to derive a model based
on the theory of phonons which are quanta of sound waves in a
macroscopic body \cite{huang}. Phonons are a mathematical analogue
of the quanta of the electromagnetic field, which are the photons,
that emerge when this field is expressed as a sum of Harmonic
oscillators. This situation is carried over to the theory of solids
which are made up of atoms that are arranged in a crystal lattice
and can be approximated by a sum of Harmonic oscillators
representing the normal modes of lattice oscillations. In this
theory, as is well known the phonons have a maximum frequency
$\omega_m$ which is given by
\begin{equation}
\omega_m = c \left(\frac{6\pi^2}{v}\right)^{1/3}\label{4e2}
\end{equation}
In (\ref{4e2}) $c$ represents the velocity of sound in the specific
case of photons, while $v = V/N$, where $V$ denotes the volume and
$N$ the number of atoms. In this model we write
$$l \equiv \left(\frac{4}{3} \pi v\right)^{1/3}$$
$l$ being the inter particle distance. Thus (\ref{4e2}) now becomes
\begin{equation}
\omega_m = c/l\label{4e3}
\end{equation}
Let us now liberate the above analysis from the immediate scenario
of atoms at lattice points and quantized sound waves due to the
Harmonic oscillations and look upon it as a general set of Harmonic
oscillators as above. Then we can see that (\ref{4e3}) and
(\ref{4e1}) are identical as
\begin{equation}
\omega = \frac{mc^2}{\hbar}\label{eZ}
\end{equation}
Using (\ref{eZ}), we can once again recover both the Planck length
and an elementary particle Compton wavelength. On the other hand it
has been shown that starting with the background Zero Point Field
the Quantum Mechanical commutation relations in $J X J$ yield the
Quantum Mechanical spin at the Compton wavelength
\cite{tduniv}. It is to be noted that this Quantum Mechanical spin feature is absent at the Planck
length.\\
We finally comment the following. If in the considerations of
Section 2 we take the particle to have negligible mass or vanishing
mass as in the case of radiation, then this new physics as embodied
in (\ref{b1}) leads to an increased frequency of the radiation, or
an apparent increase in its velocity leading to different
wavelengths travelling with different speeds. The effect is very
minute and can be observed only for very high frequency radiation
like Gamma Rays from Gamma Ray Bursts. Already there are claims that
such lags in arrival times of the Gamma Rays have indeed been found
\cite{lorentz}.

\end{document}